\documentstyle[aaspptwo,epsf]{article}

\slugcomment{Astrophysical Journal Letters, in press}

\def\mg2{Mg$_2$}

\def\ha{\relax \ifmmode {\rm H}\alpha\else H$\alpha$\fi}
\def\hb{\relax \ifmmode {\rm H}\beta\else H$\beta$\fi}
\def\hi{\relax \ifmmode {\rm H\,{\sc i}}\else H\,{\sc i}\fi}
\def\hii{\relax \ifmmode {\rm H\,{\sc ii}}\else H\,{\sc ii}\fi}
\def\h2{\relax \ifmmode {\rm H}_2\else H$_2$\fi}
\def\lha{\relax \ifmmode L_{{\rm H}\alpha}\else $L_{{\rm H}\alpha}$\fi}
\def\shi{\relax \ifmmode \sigma_{{\rm HI}}\else $\sigma_{\rm HI}$\fi}   
\def\sh2{\relax \ifmmode \sigma_{{\rm H}_2}\else $\sigma_{{\rm H}_2}$\fi}
\def\kms{\relax \ifmmode {\,\rm km\,s}^{-1}\else \,km\,s$^{-1}$\fi}

\def\min{\hbox{$^\prime$}}

\def\fdg{\hbox{$.\!\!^\circ$}}

\def\farcm{\hbox{$.\mkern-4mu^\prime$}}
\def\farcs{\hbox{$.\!\!^{\prime\prime}$}}
\def\degd#1.#2{ #1\fdg#2 }                 % degrees over decimal point     
\def\mind#1.#2{ #1\farcm#2 }               % minutes over decimal point
\def\secd#1.#2{ #1\farcs#2 }               % seconds over decimal point

\def\gtorder{\mathrel{\raise.3ex\hbox{$>$}\mkern-14mu
    \lower0.6ex\hbox{$\sim$}}}
\def\ltorder{\mathrel{\raise.3ex\hbox{$<$}\mkern-14mu
    \lower0.6ex\hbox{$\sim$}}}

\begin{document}

\title{DEFICIENCY OF `THIN' STELLAR BARS\\ IN SEYFERT HOST GALAXIES}

\author{Isaac Shlosman$^{1}$}
\author{Reynier F. Peletier$^{2}$}
\author{Johan H. Knapen$^{3,4}$}

\affil{$^{1}$ Department of Physics \& Astronomy,
University of Kentucky,
Lexington, KY 40506-0055, USA,
E-mail: {\tt shlosman@pa.uky.edu}}

\affil{$^{2}$ School of Physics and Astronomy, 
University of Nottingham, University Park,
Nottingham, NG7 2RD, UK,
E-mail: {\tt reynier.peletier@nottingham.ac.uk}}

\affil{$^{3}$Isaac Newton Group of Telescopes, Apartado 321, Santa Cruz
de La Palma, E-38700 Spain}  

\affil{$^{4}$On leave from Department of Physical Sciences,
University of Hertfordshire,
Hatfield, Herts AL10 9AB, UK,
E-mail: {\tt knapen@star.herts.ac.uk}}

\journalid{Vol}{Journ. Date}
\articleid{start page}{end page}
\paperid{manuscript id}
\cpright{type}{year}
\ccc{code}
\lefthead{}
\righthead{}

\begin{abstract}
Using all available major samples of Seyfert  galaxies
and their corresponding control samples of closely matched  non-active
galaxies, we find that the bar ellipticities (or axial ratios) in Seyfert
galaxies are systematically different from those in non-active galaxies.
Overall, there is a deficiency of bars with large ellipticities (i.e.,
`fat' or `weak' bars) in
Seyferts, compared to non-active galaxies. Accompanied with a large dispersion
due to small number statistics, 
this effect is strictly speaking at the $2\sigma$ level. 

To obtain this result, the active galaxy samples of near-infrared surface
photometry were matched to those of normal galaxies in type, host galaxy
ellipticity, absolute magnitude, and, to some extent, in redshift.
We discuss possible theoretical explanations of this phenomenon
within the framework of galactic evolution, and, in particular, of radial gas
redistribution in barred galaxies. Our conclusions provide further evidence
that Seyfert hosts differ systematically from their non-active counterparts on
scales of a few kpc.

\end{abstract}


\twocolumn

\section{Introduction}

The relationship between the large-scale morphology of Seyfert (Sy) host
galaxies and the central nonstellar activity is a long-standing problem and
the focus of an ongoing debate. Shlosman, Frank \& Begelman (1989)
argued that stellar-dynamical processes on scales of a few kpcs, and
gas-dynamical processes on smaller scales, combine to drive the gas towards
the centers and  fuel the active galactic nuclei (AGNs). Sufficient evidence,
observational  and theoretical, supports the idea that nonaxisymmetries in the
background  gravitational potential, e.g., stellar bars, induce radial mass 
redistribution in disk galaxies (e.g., Simkin, Su \& Schwarz 1980; Balick \&
Heckman 1982; Shlosman, Begelman \& Frank 1990; Athanassoula 1994; Buta \&
Combes 1996). On  the other  hand, a number of optical surveys claimed
no correlation between the large-scale  morphology and the central activity
(e.g., Moles, M\'arquez \& P\'erez 1995; Ho, Filippenko \& Sargent 1997; 
Mulchaey \& Regan 1997).  The perennial question, therefore, to be addressed
is {\it whether the AGN host galaxies differ morphologically from `normal'
(non-active) galaxies, and on what spatial scales}.  
  
High-resolution near-infrared (NIR) observations are clearly advantageous in
determining the mass distribution and hence detecting large-scale bars (McLeod
\& Rieke 1995; Mulchaey \& Regan 1997; Peletier et al. 1999, hereafter
Paper~I). Knapen, Shlosman \& Peletier (2000, Paper~II) used sub-arcsec
resolution imaging in three NIR bands ($J, H, K$) to study the complete CfA
sample of Sy's and a {\it matched} control sample of normal galaxies using
objective and stringent criteria for assigning bars. Here we look at a
different aspect, and, instead of studying the frequency of bars in galaxies,
investigate the bar axial ratios (i.e., bar ellipticities). This is done for
{\it all}   matching samples which are large enough and available in the
literature. Bar parameters, such as strength, mass, and pattern speed, are
very difficult to estimate from observations of stellar morphology alone.
Even for the simplest models, bar strength depends on bar's quadrupole moment
and on the radial distribution of axisymmetric mass in the disk, bulge and
halo. The optical RC3 catalogue (de Vaucouleurs et al. 1991) recognizes three
broad morphological classes, A, X, and B, i.e., nonbarred, intermediate barred
and strongly barred. This RC3 classification, however, is subjective and has,
to our knowledge, not been properly documented. We state that presently, it is
not feasible to estimate the distribution  of bar strengths proper in any
statistically significant sample, but bar axial ratios can provide a
reasonable alternative (Martin 1995). Here we develop this idea.

In this Letter we report a systematic difference in the distribution of the
deprojected ellipticities of large-scale stellar bars between four  
samples of Sy and normal `control' galaxies, of which three 
are independent.  We describe the samples used, provide results of our
analysis, and discuss their implications for understanding the AGN-host
galaxy connection. 

\section{Observational Database}

We use three independent samples of Sy's. NIR surface photometry is
available for two of them, allowing us to apply our criteria for bar
classification. Using a stringent criterion, we classified a galaxy barred,
(Paper~II) if {\it (i)} there is a significant rise in isophote ellipticity
followed by a significant fall, $\Delta \epsilon_{\rm gal}>0.1$, where
$\epsilon_{\rm gal}\equiv 1-b/a$ and $a$ and $b$ are semi-major and -minor
isophote axes; {\it (ii)} the position angle of isophote major axis is
constant within the bar range. The bar ellipticity was defined as
$\epsilon_{\rm b} = 10\,{\rm max}(\epsilon_{\rm gal})$ (e.g., Martin 1995).  A
galaxy is also classified as barred if the major axis position angle shows a
change of more than 75$^{\rm o}$, accompanied by an ellipticity above 0.1.  
Denoting large ellipticity (small axial ratio $b/a$)
bars with {\it Strong} (i.e., `thin' bars) and small ellipticity (large axial
ratio) bars with {\it Weak} (i.e., `fat' bars), we divided the deprojected
range of bar ellipticities $\epsilon_{\rm b}$ into two groups, $\epsilon_{\rm
b}< 4.5$ and $\geq 4.5$. Our results do not depend critically on the boundary
between these two groups (see below). Ellipticities $\epsilon_{\rm b}\leq 1$
were ignored and the galaxy was considered unbarred.  For the samples based on
the RC3 classification, `S' and `W' bar types were taken as the projected `B'
and `X' types, respectively, but due to the uncertainty in relating the RC3
morphology to even the bar ellipticities, we do {\it not} base our conclusions
on it. The following samples of Sy and matched normal (control) galaxies were
used:

\begin{enumerate}

\item {\bf KSP-RC3 Sample.} Paper~II --- CfA sample of Sy's (Huchra \& Burg
1992) observed in NIR. Morphological classification from RC3. Galaxies were
excluded when too small ($\log r_{H,19} < 0.8$, where $r_{H,19}$ is the
isophotal radius in arcsec at $H$ = 19  mag~arcsec$^{-2}$), interacting
(severely distorted or companion within 1\min), or highly inclined
($\epsilon_{\rm gal}>0.5$). A synthetic control sample of  normal galaxies
chosen from the RC3 was closely matched to CfA Sy's in morphological type
(including barred/un-barred), ellipticity and absolute magnitude (see Paper~II
for details about the technique used here).

\item {\bf MRR-RC3 Sample.} Maiolino, Ruiz \& Rieke (1995) --- an optical Sy
sample  selected from the RSA Catalogue (Sandage \& Tammann 1981). A
synthetic control sample was constructed as for the KSP-RC3. 

\item {\bf MRK-RC3 Sample.} Mulchaey, Regan \& Kundu (1997) --- subset of Maiolino 
et al. (1995) observed in NIR. Morphological classification from the RC3, control 
sample as in KSP-RC3.

\item {\bf HFS-RC3 Sample.} Ho et al. (1997) ---  very large optical sample of nearby
AGNs, from Sy's to Liners and H{\sc ii} galaxies. We have taken all galaxies
classified by Ho et al. as either Sy's or Transition objects (class T).
Morphological classification from the RC3 and control sample as in KSP-RC3. 

\item {\bf KSP Samples.} Paper~II --- projected [KSP(p)] and deprojected
[KSP(d)] bar ellipticities were determined on the basis of the photometric
analysis of our NIR images of the CfA Sy and control samples.

\item {\bf MRK Samples.} Mulchaey et al. (1997) --- projected [MRK(p)] and
deprojected [MRK(d)] bar ellipticities were determined by us from MRK's
published  ellipticities and position angle profiles for their Sy and control
samples.     

\end{enumerate}
 
\section{Statistical Results}

Since we are only concerned here with the relative distribution of S vs.
W-type bars in active and normal galaxies, we compare the frequency of
S bars for Sy's, $f_{S(Sy)}$, namely,
 \begin{equation}
 f_{S(Sy)}={[S]\over {[S+W]}}
 \end{equation}
with those in normal galaxies, $f_{S(Ctrl)}$.       
If the bar morphology in Sy's and normal galaxies is identical,
both frequencies should be the same. However,
Fig.~1 (see also Table~1) shows that in each individual sample studied there
is a systematic deficiency of S-type bars in active galaxies,  
associated, however, with a relatively large uncertainty. The overall effect is at
the $2\sigma$ level. 

\begin{figure*}[ht]
\vbox to4.6in{\rule{0pt}{4.6in}}
\includegraphics{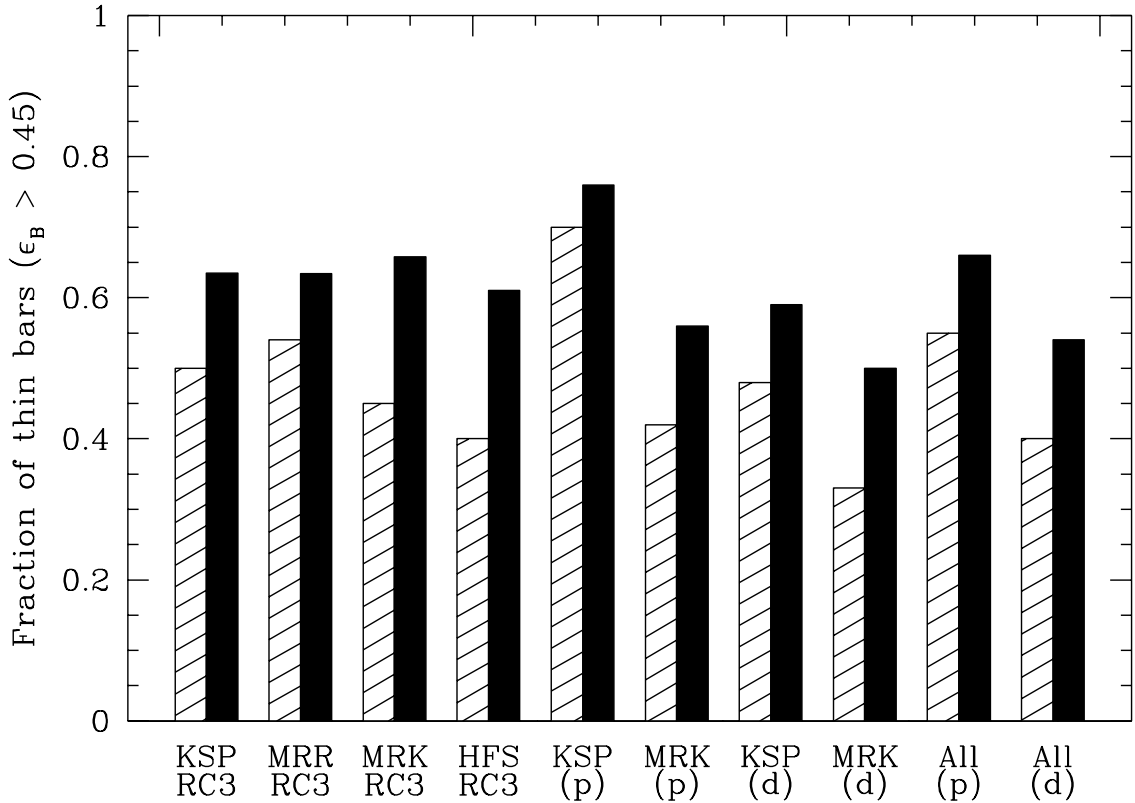}
\caption{Fraction of S-type (high-ellipticity, thin, or
`strong') bars for barred Seyferts (dashed columns) in comparison with
barred normal (control) galaxies (black columns), for all available samples.
`All' means  KSP$+$MRK. The samples appear in the order of Section~2 and
Table~1.}
\end{figure*}                           
%
%\placetable{table1} 
\begin{table}[!ht]
\begin{center}
\caption{Fractions of S- and W-type bars for different Sy and Control
samples. Tabulated are the fractions of galaxies with S-type bars,
and the number of galaxies used to obtain this number, N$_{\rm Sy}$ and
N$_{\rm Ctrl}$. 
See text for description of the samples. BX in the last column
indicates that the classification of the RC3 was used, while 
$\epsilon_{\rm b}$ means that our own ellipticity-criterion was
used.\label{table1}}  
\begin{tabular}{lccccl} \\
\hline
Sample & Sy [\%]& N$_{\rm Sy}$  & Ctrl [\%]& N$_{\rm Ctrl}$  & Crit \\
\hline
KSP-RC3 & 50$\pm$13 & 14 & 63.5$\pm$2 & $>$2000 & BX \\
MRR-RC3 & 54$\pm$7  & 54 & 63.4$\pm$2 & $>$2000 & BX \\
MRK-RC3 & 45$\pm$11 & 22 & 65.8$\pm$2 & $>$2000 & BX \\
HFS-RC3 & 40$\pm$7  & 52 & 61.0$\pm$2 & $>$2000 & BX \\
KSP(p)  & 70$\pm$10 & 23 & 76$\pm$12 & 17 & $\epsilon_{\rm b}$\\
MRK(p)  & 42$\pm$10 & 24 & 56$\pm$12 & 18 & $\epsilon_{\rm b}$\\
KSP(d)  & 48$\pm$12 & 23 & 59$\pm$12 & 17 & $\epsilon_{\rm b}$\\
MRK(d)  & 33$\pm$10 & 24 & 50$\pm$12 & 18 & $\epsilon_{\rm b}$\\  
\hline
\end{tabular}
\end{center}
\end{table}          
     
\subsection{Testing the robustness of the result}

A number of tests performed on the samples show that the result is a robust
one, showing up across all the matching samples, but at the same time
associated with relatively large uncertainties, since the available samples of
Seyferts with near-IR surface photometry are still rather small. 
We have investigated several sources of systematic errors. First, because  Sy's 
often have strong central point sources, the nuclear
PSF would make the inner regions seem rounder than they are in reality,
biasing the Sy bar shapes towards W-type bars, as compared to Control sample of
normal galaxies. We have tested this hypothesis by searching the CfA Sy's
which have a maximum in ellipticity (necessary condition for classification as
a bar; Paper~I) inside 5$''$. The effects of seeing outside this radius are
negligible when the seeing itself is smaller than  1$''$ (e.g., Peletier et
al. 1990). However, there is only one Sy galaxy, Mrk~270, which has
max($\epsilon_{\rm b}$) within this range. No bar strength could be reliably
determined for this object and it was not counted statistically, implying that
the overall distribution of bar ellipticities is not affected by the Sy nuclei.

Second, because only relative numbers are used, this result does not change if
projected or deprojected $\epsilon_{\rm b}$ are invoked for statistics. Despite
the fact that the RC3 control sample is at similar median redshift to our NIR
samples, our analysis  supports only a weak correlation between the subjective
RC3 bar classification and the bar ellipticities, in agreement with Martin
(1995, with more than 100 galaxies and RC3 classification) and Buta (1996, and 
private communications). The RC3 classifications were performed by eye, and it
is not immediately clear what exact criteria have been used for classifying a
galaxy as B or X. Out of the  40 galaxies of the KSP and control samples which
were  classified as barred by us, 16 appear unbarred (neither B nor X) in the
RC3 (Paper~II). Possible reasons for this difference include the presence of
dust, the  small size of the bar, or the inferior resolution in the
RC3. The density  contrast between the bar and the surrounding disk provides
an additional complication. When the contrast is large, it is much easier to
classify a galaxy as barred. Neither Martin's sample nor the KSP sample show
much of a correlation between $\epsilon_{\rm b}$ and B or X.  The results
based on  RC3 classifications should be, therefore, interpreted with the
necessary caution.  The strength of our analysis is in that we use both
$\epsilon_{\rm b}$ and the RC3 classification to subdivide the objects into S
and W-type bars, and both approaches supplement each other in Fig.~1 and
Table~1.

\begin{table*}[!ht]
\begin{center}
\caption{The fraction of Seyferts with thin (S) bars
for various values of the $\epsilon$-boundary dividing W and
S-type bars. `All' means KSP$+$MRK. \label{table2}}
\begin{tabular}{ccccccccccccc} \\
\hline
$\epsilon$-boundary &   KSP & Proj. &   KSP & Depr. &   MRK & Proj. &
        MRK &   Depr. & All  & Proj. & All & Depr. \\
~ & Sey. & Ctrl. &  Sey. & Ctrl. & Sey. & Ctrl. & Sey. & Ctrl. &
Sey. & Ctrl. & Sey. & Ctrl.  \\
\hline
3.5     &       0.83&0.88&0.65&0.82&0.88&0.67&0.54&0.61&0.85&0.77&0.60&0.71\\
$\pm$   &       0.08&0.08&0.10&0.09&0.07&0.11&0.10&0.12&0.05&0.07&0.07&0.08\\
4.0     &       0.83&0.88&0.52&0.77&0.63&0.67&0.50&0.56&0.72&0.77&0.51&0.66\\
$\pm$   &       0.08&0.08&0.10&0.10&0.10&0.11&0.10&0.12&0.07&0.07&0.07&0.08\\
4.5     &       0.70&0.77&0.48&0.59&0.42&0.56&0.33&0.50&0.55&0.66&0.40&0.54\\
$\pm$   &       0.10&0.10&0.10&0.12&0.10&0.12&0.10&0.12&0.07&0.08&0.07&0.08\\
5.0     &       0.70&0.77&0.44&0.53&0.38&0.44&0.17&0.50&0.53&0.60&0.30&0.51\\
$\pm$   &       0.10&0.10&0.10&0.12&0.10&0.12&0.08&0.12&0.07&0.08&0.07&0.08\\
5.5     &       0.44&0.47&0.30&0.29&0.29&0.33&0.08&0.28&0.36&0.40&0.19&0.29\\
$\pm$   &       0.10&0.12&0.10&0.11&0.09&0.11&0.06&0.11&0.07&0.08&0.06&0.08\\  
\hline
\end{tabular}
\end{center}
\end{table*}         
%\placetable{table2}
 
Third, we tested the sensitivity of our results to the assumed boundary
between S and W-type bars. This was achieved by moving this
boundary between 0.35 and 0.55 (Table 2). Moving it even more would not leave
enough galaxies in either the weak or the strong bins. The fraction  
of S or W bars does not depend critically on the
exact position of the boundary.  For each sample individually the error is
rather large, but for all the samples the frequency
of S bars in Seyferts lies below that of the control sample. To show the
significance of this result, we have constructed a larger sample by taking 
together all Seyferts of MRK and KSP, and comparing them with a Control sample
consisting of both control samples combined (the last two columns of Table~2
and Fig.~2).

\begin{figure*}[ht!!!]
\vbox to6.in{\rule{0pt}{6.in}}
\includegraphics{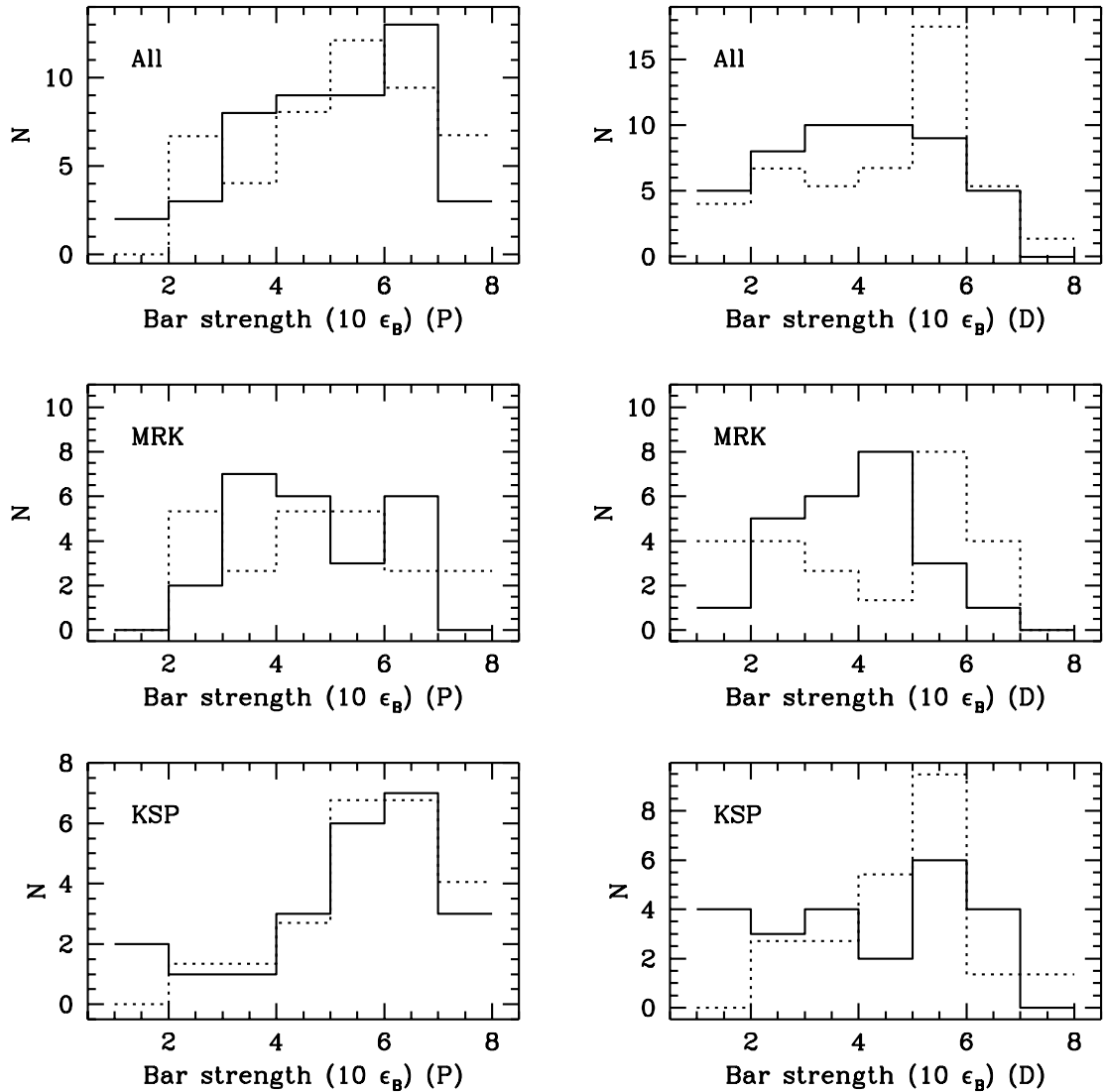}
\caption{Distribution of bar ellipticities in projected and
deprojected KSP, MRK and `All' (see the text) samples for Sy (solid lines)
and Control (dashed) galaxies. The Control distributions were scaled to Sy
ones for comparison.}
\end{figure*}  

Fourth, we have performed Kolmogorov-Smyrnov tests to check whether 
Seyfert and Control samples can be seen as drawn from the same
intrinsic  distribution. For the deprojected KSP sample the probability 
that this is in fact the case is 53\%. For the deprojected 
MRK sample we find 11\%, while for the projected samples 
the probability is 72\% and 54\%, respectively. These numbers show that this
test is inconclusive. It does not show that it is likely that the 
fractions of S and W-type bars are different in Seyferts from those in 
non-Seyferts, although it can not exclude this possibility.  

In any case, independent of the exact method by which the bar type, S or W, 
was determined (i.e., based on RC3 classification, projected or deprojected
bar ellipticities), the frequency of S bars in Seyferts is systematically 
lower than in normal galaxies.
We conclude that the result that Sy's have more W-type (`fat') bars than
S-type (`thin') bars in normal galaxies is robust, and that it is found
for a variety of samples with bars measured in a number of ways.

\section{Discussion}

All samples used in the previous section provide a coherent picture of bar
ellipticity distribution in barred Sy hosts and matching control samples of
normal barred galaxies. The most intriguing and important result is the
apparent deficiency of stellar bars with large ellipticities (i.e., large
$\epsilon_{\rm b}$) in Sy's, compared to those found in their non-active
counterparts.  Although this effect is at the level of $\sim 2\sigma$, 
due to small number statistics, it is consistent across all the Sy and
matching control samples.

There are two possible explanations for the observed phenomena within
the framework of galaxy evolution both of which are dependent on the cold gas
component in the disk. First, numerical simulations of pure stellar disks have
shown that the bar instability becomes milder, i.e., produces a bar with 
smaller ellipticities, if the disk is hotter initially, prior to instability
(Athanassoula 1983). This can be understood as weaker swing amplification in
disks with larger velocity dispersion (Toomre 1981). An additional caveat is
that the presence of a cold and clumpy gas component in the disk provides a
heating source for stars and acts to weaken {\it all} dynamical
instabilities, including the bar instability (Shlosman \& Noguchi 1993) and the
vertical bending of the bar (Berentzen et al. 1998). Gas gravity is crucial
here and even reasonable amounts of cold gas are sufficient to suppress
the instabilities altogether. If indeed Sy disks have a larger fraction of
cold gas than normal galaxies (e.g., Hunt et al. 1999), which may also be more
clumpy, this trend should be explored more fully. The resulting difference(s)
between Sy and normal disks will be long-lasting because the stellar component,
once heated up, will not be able to cool down easily, as the stellar `fluid' is
non-dissipative.  

Our understanding of the evolution of barred galaxies points to an alternative
and possibly more elegant explanation for the observed difference in bar
properties between active and normal galaxies. Numerical simulations of
bars revealed their weakening with time in response to a growing mass
concentration in the galactic centers (Hasan \& Norman 1990; Friedli \& Benz 1993; 
Hasan, Pfenniger \& Norman 1993; Berentzen et al. 1998). The radial gas inflow
towards the central kpc and further inwards is the prime suspect. All
or part of this gas can contribute to the growth of galactic bulges,
nuclear rings, disks and bars, and ultimately to central black holes (BHs)
by dissolution of the main family of periodic orbits supporting the
large-scale stellar bars. These so-called $x_1$ orbits (e.g., Sellwood \&
Wilkinson 1993) are replaced by stochastic orbits when the mass of the central
BH exceeds $\sim 1\%$ of the total mass. Heller \& Shlosman (1996) also found
that nuclear rings with masses greater than a few $\times 10^9\ {\rm M_\odot}$
lead to $x_1$ orbit dissolution exterior to the ring,  leaving smaller bar
remnants escaping detection inside the central kpc. 

In view of the fact that the supermassive BHs appear to be ubiquitous in normal
galaxies (Kormendy \& Richstone 1995) and not only in Sy nuclei, it seems
relevant to ask why the Sy hosts are affected more by the bar dissolution
processes than normal galaxies. A resolution of this paradox can lie in that
the BHs in Sy's are more massive than in their non-active counterparts, but
only statistically. If indeed supermassive BHs play a role
in the bar dissolution, we anticipate that their instantaneous mass
distributions in Sy's and normal hosts peak at different values, but have a
large overlap. Moreover, it is plausible that ground-based observations based
on stellar-dynamical considerations overestimate the BH masses in normal
galaxies due to insufficient spatial resolution. These mass estimates may
be lowered by upcoming HST observations.  

In summary, we have analyzed all available reasonably-sized 
samples of Sy host galaxies and 
independent control samples of normal galaxies  carefully matched in type,
disk ellipticity, absolute magnitude, and, to some extent, in redshift. We find
that samples of active galaxies are systematically deficient in
high-ellipticity  `thin' stellar bars compared to normal galaxies. The 
associated uncertainty
is quite large and the overall effect is at the $2\sigma$ level.  We discuss
the corollaries of this effect. The acceptable alternatives point towards the
cold and clumpy gas component in the disk as being responsible for the observed
effect. Our result provides an indication  that Sy host galaxies differ
systematically on scales of a few kpc from their normal counterparts, and that
the gas component may be responsible for this.

\acknowledgments

We thank John E.~Beckman for inviting us to the stimulating
conference during which the present paper was conceived, and Ron Buta for
helpful communications. IS research is supported in part by NASA grants
NAGW-3841, WKU-522762-98-06, HST AR-07982.01-96A and GO-08123.01-97A.

%\clearpage

%\clearpage
 

\begin{references}

\reference{} Athanassoula, E. 1983, in IAU 100 on ``Internal Kinematics \&
    Dynamics of Galaxies,'' E. Athanassoula, ed. (Dordrecht: Reidel), p.~243
\reference{} Athanassoula, E. 1994, in ``Mass-Transfer Induced Activity in
    Galaxies,'' I. Shlosman, ed. (Cambridge Univ. Press), p.~143 
\reference{} Balick, B. \& Heckman, T.M. 1982, ARAA, 20, 431 
\reference{} Berentzen, I., Heller, C.H., Shlosman, I. \& Fricke, K.    
    1998, MNRAS, 300, 49 
\reference{} Buta, R. 1996, in IAU 157 on ``Barred Galaxies,''
    R. Buta et al., eds. (San Francisco: ASP), p.~11
\reference{} Buta, R. \& Combes, F. 1996, Fund. Cosmic Phys., 17, 95
\reference{} de Vaucouleurs, G., de Vaucouleurs, A., Corwin, H.G., Buta,
    R.J., Paturel, G., Fouqu\'e, P. 1991, 3rd Reference Catalogue of Bright 
    Galaxies (RC3), Springer, New York
\reference{} Friedli, D. \& Benz, W. 1993, A\&A, 268, 65
\reference{} Hasan, H. \& Norman, C. 1990, ApJ, 361, 69
\reference{} Hasan, H., Pfenniger, D. \& Norman, C. 1993, ApJ, 409, 91
\reference{} Heller, C.H. \& Shlosman, I. 1996, ApJ, 471, 143
\reference{} Ho, L.C., Filippenko, A.V. \& Sargent, W.L.W. 1997, ApJ, 487, 591
\reference{} Huchra, J.P., \& Burg, R. 1992, ApJ, 393, 90
\reference{} Hunt, L.K., Malkan, M.A., Moriondo, G. \& Salvati, M. 1999, ApJ,
    510, 637 
\reference{} Knapen, J.H., Shlosman, I. \& Peletier, R.F. 2000, ApJ, 529,
   93 ({\bf Paper~II})
\reference{} Kormendy, J. \& Richstone, D. 1995, ARAA, 33, 581 
\reference{} McLeod, K.K. \& Rieke, G.H. 1995, ApJ, 441, 96 
\reference{} Maiolino, R., Ruiz, M. \& Rieke, G.H. 1995, ApJ, 446, 561
\reference{} Martin, P. 1995, AJ, 109, 2428
\reference{} Moles, M., Marquez, I. \& Perez, E. 1995, ApJ, 438, 604
\reference{} Mulchaey, J. \& Regan, M. 1997, ApJ, 482, L135
\reference{} Mulchaey, J., Regan, M. \& Kundu, A. 1997, ApJS, 110, 299
\reference{} Peletier, R.F., Davies, R.L., Illingworth, G., Davis, L. \& 
   Cawson, M. 1990, AJ, 100, 1091
\reference{} Peletier, R.F., Knapen, J.H., Shlosman, I.,
   P\'erez-Ramirez, D., Nadeau, D., Doyon, R., Rodriguez-Espinosa, J.M. \&
   P\'erez-Garc\'\i a, A.M. 1999, ApJS, 125, 363 ({\bf Paper~I})
\reference{} Sandage, A. \& Tammann, G. 1981, The Revised Shapley-Ames
   Catalogue, Carnegie Inst. of Washington, Washington DC
\reference{} Sellwood, J.A. \& Wilkinson, A. 1993, Rep. Prog. Phys., 56, 173
%\reference{} Shlosman, I. 1996, in Proc. Nobel Symp. on ``Barred Galaxies \& 
%   Circumnuclear Activity,'' A. Sandqvist \& P.O. Lindblad, eds.
%   (Springer-Verlag), p.~141  
\reference{} Shlosman, I., Begelman, M.C. \& Frank, J. 1990, Nature, 345, 679
\reference{} Shlosman, I., Frank, J. \& Begelman, M.C. 1989, Nature, 338, 45 
\reference{} Shlosman, I. \& Noguchi, M. 1993, ApJ, 414, 474
\reference{} Simkin, S.M., Su, H.J. \& Schwarz, M.P. 1980, ApJ, 237, 404
\reference{} Toomre, A. 1981, in ``Structure \& Evolution of Normal
   Galaxies,'' S.M. Fall \& D. Lynden-Bell, eds. (Cambridge Univ. Press),
   p.~111

\end{references}
\end{document}